\documentclass[12pt]{article}
\begin{document}

\begin{center}
\textbf{O.A.Khrustalev M.V.Tchitchikina }

\textbf{QUANTUM GRAVITY on the }

\textbf{CLASSICAL BACKGROUND: GROUP ANALYSIS, part II.}
\end{center}

\section{Introduction}

{\normalsize In the article }$\left[ 28\right] ${\normalsize we have
represented an approach to the description of gravitational field from the
point of view of quantum field theory. We consider quantization in the
neighbourhood of solution of Einstein equation and use the method of
Bogoliubov transformations for this purpose. }We applied Bogoliubov
transformation to the quantization of gravitational field in the
neighbourhood of nontrivial classical component, that permitted us to avoid
zero-mode problem. Einstein equations for the classical component has been
obtained as a necessary condition for the perturbation theory to be
applicable.{\normalsize \ We would like to underline that Einstein equations
has been obtained in the process of perturbation theory construction as a
condition of validity, not as a sequence of variational principle.}

The expression for quantum corrections of the field operator and explicit
view of state is the task of the present article.

{\normalsize \ }

\section{System State Space construction.}

{\normalsize Now the task is to find the expressions for the action with
second order accuracy with respect to inverted power of coupling constant.
For that purpose it is necessary to construct system state space, in which
one can make reduction of field state number and finds expressions for
action derivatives with respect to symmetry group generators. }

{\normalsize The action in the next order represent at the Appendix2. }

{\normalsize Using equations (I.8) that has been obtained in the }$\left[
28\right] $ for the classical part {\normalsize \ we can state that
Hamiltonian looks like: }

$H${\normalsize $_2=-in^lA_l^p(a)\frac \partial {\partial
a^p}+\int\limits_\Sigma aH_2(\hat{P},\hat{Q})+$}

{\normalsize $i\int\limits_\Sigma u_{stn}\frac \delta {\delta
u_{st}}+u_{nn}^{st}\frac \delta {\delta u_n^{st}}+r_k\int\limits_\Sigma
N_{nn_{}}^{stk}u_{st}-N_{stn}^ku_n^{st}.$}

{\normalsize The number of independent variables has been doubled due to one
consider $f_{st}(x)$ and $f_n^{st}(x)$ as independent. Cause of additional
condition (3), which connect $u_{st}(x^{^{\prime }})$ and $%
u_n^{st}(x^{^{\prime }})$, the number of independent variables (minus group
variables) become equal $(2*\infty -r)$. To reduce them to $(\infty -r)$,
one needs also }$r$ {\normalsize conditions (Remind that }$r$ {\normalsize %
is the number of group parameters). }

{\normalsize For that end let's represent $u_{st}(x^{^{\prime }})$ in the
form: }

{\normalsize $u_{st}(x^{\prime })=w_{st}(x^{\prime })+N_{st}^a(x^{\prime
})r_a,$ $u_n^{st}(x^{\prime })=w_n^{st}(x^{\prime })+N_n^{st^a}(x^{\prime
})r_a\backslash eqno(1).$}

{\normalsize We consider those relationships to express $u_{st}(x^{^{\prime
}})$, $u_n^{st}(x^{^{\prime }})$ in the terms of independent variables $r_a$
and new functions $w_{st}(x^{^{\prime }})$, $u_n^{st}(x^{^{\prime }})$. It
is possible in the case when $(w_{st}(x^{^{\prime }}),w_n^{st}(x^{^{\prime
}}))$ are connected by }$r${\normalsize \ linear relationships. If $%
N^a(x^{^{\prime }})$ will be chosen so that the following conditions are
performed: }

{\normalsize $\omega (N^{sta},N^{stb})=0,$ }

{\normalsize then those conditions could be formulated as six conditions,
that $w_{st}(x^{^{\prime }})$, $w_{st}(x^{^{\prime }})$ satisfy: }

{\normalsize $\omega (N^{sta},w_{st})=0,$ $\omega (M_{sta},w_{st})=0.$}

{\normalsize Operators $\hat{Q}$, $\hat{P}$ in the terms of new variables
reads: }

{\normalsize $\hat{Q}_{st}(x^{\prime })=\hat{Q}_{st}(x^{\prime
})+q_{st}(x^{\prime }),$ $\hat{P}^{st}(x^{\prime })=\hat{P}^{st}(x^{\prime
})+p^{st}(x^{\prime }),$ }

{\normalsize where }

$\bar{Q}${\normalsize $_{st}(x^{\prime })=\frac 1{\sqrt{2}}\left(
w_{st}(x^{\prime })+i\frac \delta {\delta w_n^{st}(x^{\prime })}\right) ,%
\bar{P}^{st}(x^{\prime })=\frac 1{\sqrt{2}}\left( w_n^{st}(x^{\prime
})-i\frac \delta {\delta w_{st}(x^{\prime })}\right) ,$ }

{\normalsize and }

{\normalsize $q_{st}(x^{\prime })=\frac i{\sqrt{2}}\frac{\delta r_a}{\delta
u_n^{st}(x^{\prime })}\frac \partial {\partial r_a},$ $p^{st}(x^{\prime })=%
\frac{-i}{\sqrt{2}}\frac{\delta r_a}{\delta u_{st}(x^{\prime })}\frac
\partial {\partial r_a}.$ }

{\normalsize Necessary reduction of the states number can be made by the
following way: let's suppose that field condition is defined by functionals
of $w_{st}(x^{^{\prime }})$ and $w_n(x^{^{\prime }})$, in which $\bar{Q}%
(x^{^{\prime }})$ and $\bar{P}(x^{^{\prime }})$ become following ones:}

{\normalsize $\bar{Q}(x^{^{\prime }})\longrightarrow {\frac 1{\sqrt{2}}}%
w_{st}(x^{^{\prime }}),\qquad \bar{P}(x^{^{\prime }})\longrightarrow i{\frac
1{\sqrt{2}}}{\frac \delta {\delta w_{st}(x^{^{\prime }})}}.$\ }

{\normalsize The variables $r_a$ have not physical sense. They have appeared
as a rest of the state space reduction in the terms of Bogoliubov group
variables. We will show that separation of these variables is connected with
integrals of motion structure in the zero-point order, so it is dynamic by
nature. }

{\normalsize The example of reduction will be given below on the example of
translation group for the simplicity. }

{\normalsize \ }

\section{\protect\normalsize Translation group }

{\normalsize In the case if symmetry group is the group of space-time
translations }

{\normalsize $x^{\prime \alpha }=x^\alpha -\tau ^\alpha ,$ $\left( \delta
x^{\prime }\right) ^\alpha =\delta \tau ^\alpha .$ }

{\normalsize we have the same expressions for the operators of coordinate
and momentum of field operators: }

{\normalsize $\hat{q}_{st}=g\left( F_{st}(x^{\prime })+\frac 1g\hat{Q}%
_{st}(x^{\prime })+\frac 1{g^2}A_{st}(x^{\prime })\right) ,$ }

{\normalsize $\hat{p}^{st}=g\left( F_n^{st}(x^{\prime })+\frac 1g\hat{P}%
^{st}(x^{\prime })+\frac 1{g^2}A_n^{st}(x^{\prime })\right) .$ }

(see Appendix 2)

{\normalsize As well as in the general case the necessary condition of
perturbation theory to be applicable is the Einstein equation performance.
And the second order with respect to coupling constant reads:\ }

$H${\normalsize $_2=-in^p\frac \partial {\partial \tau
^p}+\int\limits_\Sigma aH_2(\hat{P},\hat{Q})+$}

{\normalsize $\int\limits_\Sigma u_{stn}\frac \delta {\delta
u_{st}}+u_{nn}^{st}\frac \delta {\delta u_n^{st}}+r_k\int\limits_\Sigma
N_{nn_{}}^{stk}u_{st}-N_{stn}^ku_n^{st},$}

where

{\normalsize $H_2(\hat{P},\hat{Q})=\frac 1{\sqrt{F}}$}$\left( \left(
2F_n^{st}F_{st}+\frac 12F_n\right) \hat{P}^{st}\hat{Q}_{st}+\frac
12F_n^{}F_{}^{st}F_n^{st}\hat{Q}_{st}\hat{Q}_{st}\right) +$

{\normalsize $\frac 1{\sqrt{F}}$}$\left( \hat{P}^{st}\hat{P}_{st}-\frac 12%
\hat{P}_{}^2\right) ${\normalsize $-\sqrt{F}\left( 
\begin{array}{c}
\left( R^{st}\hat{Q}^{st}\hat{Q}_{st}-\frac 12R^{st}F^{st}\hat{Q}_{st}\hat{Q}%
_{st}\right) - \\ 
\hat{Q}^{st}R_{st}(\hat{Q})-\frac 12R(\hat{Q})F^{st}\hat{Q}_{st}
\end{array}
\right) $ }

{\normalsize The surface $\Sigma $ is assumed to be plane and defined via 3
ortonormal vectors $\vec{e}_a:$ }

{\normalsize $e_a^\alpha e_{b\alpha }^{}=-\delta _{ab},$ $e_\beta
^ae_{a\alpha }^{}=g_{\alpha \beta },$ }

{\normalsize and normal vector $\vec{n}:$ }

{\normalsize $\vec{e}_a^{}\vec{n}=0,$ $\vec{n}^2=1,$ }

{\normalsize normal derivative is given by the formula }

{\normalsize $\frac \partial {\partial n}=n^\alpha \frac \partial {\partial
x^\alpha }.$ }

{\normalsize For simplicity we consider the plane $t=0$ and hence }

{\normalsize $\frac \partial {\partial n}=\frac \partial {\partial t}.$ }

{\normalsize Coordinates $x^{\prime \alpha }$ are defined with a help of
parameters $\lambda ^a:$ }

{\normalsize $x^{\prime \alpha }=e_a^\alpha \lambda ^a,$ $a=1,2,3,$ $\alpha
=0,1,2,3.$ }

{\normalsize Without restriction of generality we can consider the case when 
$F_{st}(x^{^{\prime }})$ satisfy to the boundary condition on $\Sigma $: }

{\normalsize $F_n^{st}(x^{^{\prime }})=0.$}

{\normalsize In that case }

{\normalsize $M_{\alpha st}(x^{\prime })=-\frac 1{\sqrt{2}}e_a^\alpha
F_{st\lambda _a}(x^{\prime }),$ $M_{n\alpha }^{st}(x^{\prime })=\frac 1{%
\sqrt{2}}n_\alpha F_{nn}^{st}(x^{\prime }).$ }

{\normalsize Necessary conditions }

{\normalsize $\omega \left( N_{st}^a,M_b^{st}\right) =\delta _b^a,$ $\omega
\left( N^{ast},N_{st}^b\right) =0,$ }

{\normalsize are performed if $\tilde{N}^\alpha (x^{^{\prime }})$ are
defined by formulae: }

{\normalsize $N_{st}^\alpha (x^{\prime })=An^\alpha F_{stnn}^{}(x^{\prime
}), $ $N_n^{\alpha st}(x^{\prime })=B^{ab}e_a^\alpha F_{\lambda
_b}^{st}(x^{\prime }),$ }

{\normalsize where }

{\normalsize $A=-\frac{\sqrt{2}}{\int\limits_\Sigma F_{stnn}^2(x^{\prime })}%
, $ $B^{ab}=\frac{\sqrt{2}}{\int\limits_\Sigma F_{\lambda _a}^{st}(x^{\prime
})F_{st\lambda _b}^{}(x^{\prime })}.$ }

{\normalsize Conditions which must be satisfied by functions $%
w_{st}(x^{^{\prime }})$, $w_n^{st}(x^{^{\prime }})$ show us that $%
w_{st}(x^{^{\prime }})$ and $w_n^{st}(x^{^{\prime }})$ can be obtained from
independent $f_{st}(x)$ and $f_n^{st}(x)$ by the same self-conjugated
projection. It permits us, for example, convert to variables $%
z_{st}(x^{^{\prime }})$ and $z^{^{st}*}(x^{^{\prime }})$ and holomorphic
representation connected with them.\ }

\section{Creation-Annihilation Operators:}

{\normalsize After representation of quantum addend in the form (1) the
action at the second order with respect to $g$ reads:}

{\normalsize \ $H_2=in^\alpha \frac \partial {\partial \tau ^\alpha
}+H_{21}+H_{22},$}

here

$H${\normalsize $_{21}=\int\limits_\Sigma \left( -\frac 1{2\sqrt{F}}\left(
\frac \delta {\delta w_{st}}\frac \delta {\delta w^{st}}-\frac 12\frac{%
\delta ^2}{\delta w^2}\right) -\sqrt{F}\left( Rw^{st}-R^{st}(w)\right)
w_{st}\right) +$ }

{\normalsize $+i\int\limits_\Sigma w_{stn}\frac \delta {\delta
w_{st}}+w_{nn}^{st}\frac \delta {\delta w_n^{st}},$ }

{\normalsize $H_{22}=\int\limits_\Sigma \left( 
\begin{array}{c}
-\frac 1{2\sqrt{F}}\left( \frac{\delta r_a}{\delta u_{st}}\frac{\delta r_b}{%
\delta u^{st}}-\frac 12F^{st}\delta _{ab}\frac{\delta r_a}{\delta u}\frac{%
\delta r_b}{\delta u_{st}}\right) + \\ 
+\frac{\sqrt{F}}4\left( R\frac{\delta r_a}{\delta u_n^{st}}-R^{st}\left( 
\frac{\delta r_a}{\delta u_n^{st}}\right) \right) \frac{\delta r_a}{\delta
u_{nst}}
\end{array}
\right) \frac{\partial ^2}{\partial r_a\partial r_b}+$ }

{\normalsize $r_a\frac \partial {\partial r_b}\left( i\int\limits_\Sigma
N_{nn_{}}^{st^a}\frac{\delta r_b}{\delta u_{st}}-N_{stn}^a\frac{\delta r_b}{%
\delta u_n^{st}}\right) -r_ar_b\left( \int\limits_\Sigma
N_{nn_{}}^{st^a}N_{st}^b-N_{stn}^bN_{n_{}}^{st^a}\right) ,$ \ }

{\normalsize $A=\frac{\delta u_n^{st}}{\delta r_a}R_{st}\left( \frac{\delta
r_a}{\delta u_n^{}}\right) .$ }

{\normalsize The operators $H_2$ act at the space like $F[w,w_n]F[r],$ so $%
H_{21}$ act at the space $F[w,w_n]$ but operator $H_{22}$ act at the space $%
F[r].$ Those spaces are orthohonal. Representation of $w_{st}(x)$ as a
following series is correct at least at the order $O(t^2)$:\ }

{\normalsize $w_{st}(x)=\sum\limits_m\phi _{stm}(\vec{x})c_m(t)$ $\frac
\delta {\delta w_{st}}(x)=\sum\limits_m\phi _{stm}(\vec{x})\frac \delta
{\delta c_m(t)},$ }

{\normalsize where $\phi _{st_m}(x)$ are orthonormal functions }

{\normalsize $\int\limits_\Sigma \phi _{stm}(\vec{x})\phi _{_n}^{st}(\vec{x}%
)=\delta _{mn}.$}

{\normalsize They are the solutions of the equations: }

{\normalsize $\phi _{stm}(\vec{x})=\frac{2a}{\sqrt{F}}\left( \phi _{stm}(%
\vec{x})-\frac 12F_{st}\phi _m(\vec{x})\right) ,$ }

{\normalsize $\omega _m^2\phi _m^{st}(\vec{x})=a\sqrt{F}\left( R\phi _m^{st}(%
\vec{x})-R^{st}(\phi _m(\vec{x}))\right) .$ }

{\normalsize Creation-annihilation operators can be represented as following
functions: }

{\normalsize $a_s=\sqrt{\Omega _s}c_s+\frac 1{2\sqrt{\Omega _s}}\frac
\partial {\partial c_s},$ $a_s^{\dagger }=\sqrt{\Omega _s}c_s-\frac 1{2\sqrt{%
\Omega _s}}\frac \partial {\partial c_s}.$ }

{\normalsize The functions $w_{st}(x^{^{\prime }})$, being given as series
by $\phi _{st_m}(x)$, satisfy to the wave equation (at least at the small }$%
t ${\normalsize ), if dependence of $a$ and $a^{\dagger }$ on time is
determined by the Heizenberg equations: }

{\normalsize $a_s(t)=e^{-i\Omega _st}a_s,$ $a_s^{\dagger }(t)=e^{i\Omega
_st}a_s^{\dagger }.$ }

{\normalsize Dependence of $w_{st}$ on $t$ looks like:}

{\normalsize \ $w_{st}(t)=\sum\limits_m\frac 1{2\sqrt{\Omega _m}}\phi _{stm}(%
\vec{x})\left( a_me^{-i\Omega _mt}+a_m^{\dagger }e^{i\Omega _mt}\right) ,$ }

{\normalsize $w_{t_{st}}(t)=-i\sum\limits_m\frac{\sqrt{\Omega _m}}2\phi
_{stm}(\vec{x})\left( a_me^{-i\Omega _mt}-a_m^{\dagger }e^{i\Omega
_mt}\right) .$}

{\normalsize and it shows that}

{\normalsize $H_{21}=-\frac 14\sum\limits_m\left( \frac \delta {\delta
c_m}\right) ^2+\omega _m^2c_m^2,$}

where{\normalsize \ }

{\normalsize And $H_2$ looks like:}

{\normalsize $H_2=in^\alpha \frac \partial {\partial \tau ^\alpha }+H_{22},$}

{\normalsize Operators $H_{22}$ contain only exceed variables $r_a$, ${\frac
\partial {\partial r_a}}$ . }

{\normalsize Let's denote: }

{\normalsize $r_{a_{\Vert }}=e_a^\alpha r_\alpha ^{},$ $r_{\bot }=n^\alpha
r_\alpha ^{},$ $\frac \partial {\partial r_{a_{\Vert }}}=e_a^\alpha \frac
\partial {\partial r_\alpha },$ $\frac \partial {\partial r_{\bot
}}=n^\alpha \frac \partial {\partial r_\alpha },$ }

{\normalsize and recollect that \ }

{\normalsize $\frac{\delta r_a}{\delta u_{st}(x^{\prime })}=-\frac 1{\sqrt{2}%
}n_\alpha F_n^{st}(x^{\prime }),$ $\frac{\delta r_a}{\delta
u_{nn}^{st}(x^{\prime })}=-\frac 1{\sqrt{2}}e_\alpha ^aF_{\lambda
_a}^{st}(x^{\prime }).$ }

{\normalsize If we use the following definitions: }

{\normalsize $a^{ab}=\frac 2{\int\limits_\Sigma F_{\lambda
_a}^{st}(x^{\prime })F_{st\lambda _a}^{}(x^{\prime })},$ $b=\frac
2{\int\limits_\Sigma F_{_{nn}}^{st}(x^{\prime })F_{st}^{}(x^{\prime })},$ }

{\normalsize $c^{ab}=\frac 12\int\limits_\Sigma \frac{\sqrt{F(x^{\prime })}}%
4\left( RF_{\lambda _a}^{st}(x^{\prime })-R^{st}\left( F_{\lambda _a}\right)
\right) F_{st\lambda _b}^{}(x^{\prime }),$ }

{\normalsize $d=-\frac 12\int\limits_\Sigma \frac{F_{st_{nn}}^{}(x^{\prime })%
}{2\sqrt{F}}\left( F_{_{nn}}^{st}(x^{\prime })-\frac 12F^{st}(x^{\prime
})F_{_{nn}}(x^{\prime })\right) ,$ }

{\normalsize than we have }

{\normalsize $H_{22}=a^{ab}r_{a_{\Vert }}r_{b_{\Vert }}+br_{\bot }^2+c^{ab}%
\frac{\partial ^2}{\partial r_a\partial r_b}+d\frac{\partial ^2}{\partial
r_{\bot }^2}.$ }

{\normalsize On the following state vector $S_{22}$ is equal zero: }

{\normalsize $\Psi =\psi e^{\alpha r_{a_{\Vert }}^2+\beta r_{b_{\Vert
}}^2+\gamma r_{\bot }^2}$ }

{\normalsize where parameters satisfy equations: }

{\normalsize $a^{ab}+4\alpha \beta c^{ab}=0,$ }

{\normalsize $b+4\gamma ^2d=0.$ }

\section{Conclusion}

{\normalsize After removing of exceed variables action in the zero-point
order looks like:}

{\normalsize $H_{22}=-i\frac \partial {\partial t}$}

{\normalsize Analogously it is possible to show that in the case of general
group }

{\normalsize $H_{03}=-in^lA_l^p(a)\frac \partial {\partial a^p}.$ }

{\normalsize Field operator $\psi (x)$ looks like:}

{\normalsize \ $\psi _{st}(x)=gF_{st}(x^{^{\prime }})+\hat{\Phi}%
_{st}(x^{^{\prime }})+\hat{\phi}_{st_a}{\frac \partial {\partial r_a}}+{%
\frac 1g}A_{st}(x^{^{\prime }},\tau ),$}

{\normalsize here $\hat{\Phi}_{st}(x^{^{\prime }})$ is the solution of the
evolution equation: }

{\normalsize $\hat{\Phi}_{{st}_n}=\frac{2a}{\sqrt{\hat{\Phi}}}\left( \hat{%
\Phi}_{nst}-\frac 12\hat{\Phi}_n\hat{\Phi}_{st}\right) ,$ }

{\normalsize $\hat{\Phi}_{nn}^{st}=\frac a{2\sqrt{\hat{\Phi}}}\left( \hat{%
\Phi}_{nkl}\hat{\Phi}_n^{kl}-\frac 12\hat{\Phi}_n^2\right) \hat{\Phi}^{st}-%
\frac{2a}{\sqrt{\hat{\Phi}}}\left( \hat{\Phi}_n^{st}\hat{\Phi}_n^{kl}\hat{%
\Phi}_{nkl}^{}-\frac 12\hat{\Phi}_n\hat{\Phi}_n^{st}\right) -$ }

{\normalsize $-a\sqrt{\hat{\Phi}}\left( R^{st}-\frac 12\hat{\Phi}%
_{}^{st}R\right) -\sqrt{\hat{\Phi}}\left( \hat{\Phi}^{sl}c_{;l}^t-\hat{\Phi}%
^{st}c_{;l}^l\right) ,$ }

{\normalsize with a boundary condition on the $\Sigma $: 
\[
\hat{\Phi}_{st}=\hat{Q}_{st}(x^{^{\prime }}),\qquad \hat{\Phi}_t^{st}=\hat{P}%
^{st}(x^{^{\prime }}). 
\]
}

We applied Bogoliubov transformation to the quantization of gravitational
field in the neighbourhood of nontrivial classical component, that permitted
us to avoid zero-mode problem.

Einstein equations for the classical component has been obtained as a
necessary condition for the perturbation theory to be applicable, not as a
sequence of variational principle.

We obtained expression for quantum corrections of the field operator and
explicit view of state vector, that permits us to calculate quantum
corrections to the observables like effective mass, energy spectrum and so
on.

Such kind of calculations for the physically interesting cases like Kerr,
Schwartzshild and others exact solutions of Einstein equation are the
nearest future project research; however those calculations demands only
high level mathematical techniques, while the main principles of our
approach are represented in the paper above.

\section{\protect\normalsize Appendix 1. }

{\normalsize Let's consider expansion of Hamiltonian into the series with
respect to inverted powers of coupling constant: }

{\normalsize $H=\frac 1{\sqrt{\gamma }}\left( \pi _{st}\pi ^{st}-\frac 12\pi
^2\right) -\sqrt{\gamma }R$ }

{\normalsize \ the 2dn order with respect to inverted powers of coupling
constant: }

{\normalsize $H_2=\frac 1{\sqrt{F}}\left( 
\begin{array}{c}
-\frac 12\left( F_{n_{kl}}F_n^{kl}-\frac 12F_n^2\right) F^{st}A_{st}+ \\ 
\left( A_n^{st}F_{n_{st}}+F_n^{st}D_{st}+\hat{P}^{st}S_{st}\right) - \\ 
F_n\left( A_n+A_{st}F_n^{st}+\hat{Q}_{st}\hat{P}^{st}\right) -\frac 12F_n(%
\hat{P}+F_n^{st}\hat{Q}_{st})^2 \\ 
\frac 12F^{st}\hat{Q}_{st}\left( \left( \hat{P}%
^{st}F_{n_{st}}+S_{st}F_n^{st}\right) -F_n(\hat{P}+F_n^{st}\hat{Q}%
_{st})\right)
\end{array}
\right) -$ }

{\normalsize $-\sqrt{F}\left( 
\begin{array}{c}
\frac 12RF_{}^{st}A_{st}+R(F,\hat{Q},A)+B_{}^{st}R_{st}-\hat{Q}%
_{}^{st}R_{st}(F,\hat{Q}) \\ 
+\frac 12\left( F_{}^{st}R(F,\hat{Q})-\hat{Q}_{}^{st}R_{st}\right) F_{}^{st}%
\hat{Q}_{st}
\end{array}
\right) .$ }

{\normalsize Lets' consider the addend: }

{\normalsize $a\sqrt{F}F_{}^{st}F_{}^{st}\hat{Q}_{st}R_{st}(F,\hat{Q})=$ }

{\normalsize $Div+\sqrt{F}F^{st}\left( a\hat{Q}\right) _{;t}\Gamma _{sl}^l(%
\hat{Q})-\sqrt{F}F^{st}\left( a\hat{Q}\right) _{;l}\Gamma _{st}^l(\hat{Q}).$ 
}

{\normalsize Denote }

{\normalsize $r^s=\left( a\hat{Q}\right) _{;t}F^{st},$ }

{\normalsize so straightforward calculations (see Appendix1 [28]) shows us
that }

{\normalsize $a\sqrt{F}F_{}^{st}\hat{Q}R_{st}(F,\hat{Q})=\sqrt{F}\hat{Q}%
\left( F^{sp}r_{;p}^t-F^{st}r_{;p}^p\right) =$ }

{\normalsize $\sqrt{F}\left( F^{sp}c_{;p}^t-F^{st}c_{;p}^p\right) \hat{Q}%
_{st}\hat{Q}_{st}F^{st}+a\sqrt{F}F^{st}\left(
F^{sp}F^{tr}-F^{st}F^{pr}\right) \hat{Q}_{st,_{pr}},$}

{\normalsize and we can state that the addend is looks like the following: }

{\normalsize $a\sqrt{F}\hat{Q}_{}^{st}R_{st}(F,\hat{Q})=\sqrt{F}\left( a\hat{%
Q}^{st}\right) _{;t}\Gamma _{sl}^l(\hat{Q})-\sqrt{F}\left( a\hat{Q}%
^{st}\right) _{;l}\Gamma _{st}^l(\hat{Q})=$ }

{\normalsize $\sqrt{F}\left( F^{sp}a_{;tp}\hat{Q}^{st}-F^{st}a_{;tp}\hat{Q}%
^{pt}\right) \hat{Q}_{st}+a\sqrt{F}F^{st}\left( F^{sp}\hat{Q}%
_{;tp}^{st}-F^{st}\hat{Q}_{;tp}^{pt}\right) \hat{Q}_{st}.$}

so we can satae that

{\normalsize $H_2=\int\limits_\Sigma \left(
F_{n_{st}}A_n^{st}+D_{st}F_n^{st}+\hat{Q}_{n_{st}}\hat{P}^{st}+aH_2\right) ,$
}

{\normalsize here }

{\normalsize $aH_2=aH_2(A)+aH_2(\hat{P},\hat{Q})$ }

{\normalsize and }

{\normalsize $aH_2(A)=$ $\frac{2a}{\sqrt{F}}\left( F_{n_{st}}-\frac
12F_nF_{st}\right) A_n^{st}-$}

{\normalsize $-\frac a{2\sqrt{F}}\left( F_{n_{kl}}F_n^{kl}-\frac
12F_n^2\right) F^{st}A_{st}+\frac{2a}{\sqrt{F}}\left(
F_n^{st}F_n^{kl}F_{n_{st}}-\frac 12F_nF_n^{st}\right) A_{st}$ }

{\normalsize $+a\sqrt{F}\left( R^{st}-\frac 12F_{}^{st}R\right) A_{st}+\sqrt{%
F}\left( F^{sl}c_{;l}^t-F^{st}c_{;l}^l\right) A_{st},$}

{\normalsize $H_2(\hat{P},\hat{Q})=\frac 1{\sqrt{F}}\left( 
\begin{array}{c}
2F_n^{st}F_{st}\hat{P}^{st}\hat{Q}_{st}+F_n^{st}F_n^{st}\hat{Q}_{st}\hat{Q}%
_{st}+\hat{P}^{st}\hat{P}_{st}+ \\ 
+2F_n^{st}F_{st}\hat{P}^{st}\hat{Q}_{st}-\frac 12\hat{P}_{}^2-F_n^{st}F_{st}%
\hat{P}^{st}\hat{Q}_{st}- \\ 
-\frac 12F_n^{st}F_n^{st}\hat{Q}_{st}\hat{Q}_{st}-\frac 12F_n^{st}F_{st}\hat{%
P}^{st}\hat{Q}_{st}- \\ 
-\frac 12F_n^{st}F_{st}\hat{P}^{st}\hat{Q}_{st}-\frac 12F_n^{st}F_n^{st}\hat{%
Q}_{st}\hat{Q}_{st}+ \\ 
+\frac 12F_n\hat{P}^{st}\hat{Q}_{st}+\frac 12F_n^{}F_{}^{st}F_n^{st}\hat{Q}%
_{st}\hat{Q}_{st}
\end{array}
\right) $ }

\section{{{\protect\normalsize $-\protect\sqrt{F}\left( \left( R^{st}\hat{Q}%
^{st}\hat{Q}_{st}-\frac 12R^{st}F^{st}\hat{Q}_{st}\hat{Q}_{st}\right) -\hat{Q%
}^{st}R_{st}(\hat{Q})-\frac 12R(\hat{Q})F^{st}\hat{Q}_{st}\right) .$}}}

\section{Appendix2}

{\normalsize Translation group }

{\normalsize In the case if symmetry group is the group of space-time
translations }

{\normalsize $x^{\prime \alpha }=x^\alpha -\tau ^\alpha ,$ $\left( \delta
x^{\prime }\right) ^\alpha =\delta \tau ^\alpha .$ }

{\normalsize the number of group parameters is equal 4 and Bogoliubov
transformation reads as usual: }

{\normalsize $f_{st}(x)=gv_{st}(x^{\prime })+u_{st}(x^{\prime })$, $%
f_n^{st}(x)=gv_n^{st}(x^{\prime })+u_n^{st}(x^{\prime }),$ }

{\normalsize and additional conditions are:\ }

{\normalsize $\int\limits_\Sigma d\sigma \left( N_n^{st^k}(x^{\prime
})u_{st}(x^{\prime })-N_{st}^k(x^{\prime })u_n^{st}(x^{\prime })\right) =0.$ 
}

{\normalsize In this case we define $M_{st_r}(x^{\prime })$ as }

{\normalsize $v_{st_r}(x^{\prime })=-M_{st_r}(x^{\prime }),$ $%
v_{n_r}^{st}(x^{\prime })=-M_{n_r}^{st}(x^{\prime }).$ }

{\normalsize Equations that express group generators as functionals of field
functions are: }

{\normalsize $\frac{\delta \tau ^p}{\delta f_{st}(x)}=\frac
1gT_k^pN_n^{st^k}(x^{\prime }),$ $\frac{\delta \tau ^p}{\delta f_n^{st}(x)}%
=\frac 1gT_k^pN_{st}^k(x^{\prime }),$ $T_s^l=\delta _s^l-\frac 1gT_s^rR_r^l,$
}

{\normalsize and after canonical transformation (6) }

{\normalsize $-i\frac \partial {\partial \tau ^p}\longrightarrow
g^2J_p-i\frac \partial {\partial \tau ^p}$ }

{\normalsize and definitions }

{\normalsize $F_{st}(x^{\prime })=\frac 1{\sqrt{2}}\left( v_{st}(x^{\prime
})+N_{st}^k(x^{\prime })J_k\right) ,$ }

{\normalsize $F_n^{st}(x^{\prime })=\frac 1{\sqrt{2}}\left(
v_n^{st}(x^{\prime })+N_n^{st^k}(x^{\prime })J_k\right) ,$ }

{\normalsize $\hat{Q}_{st}(x^{\prime })=\frac 1{\sqrt{2}}\left(
u_{st}(x^{\prime })+i\frac \delta {\delta u_n^{st}(x^{\prime
})}-N_{st}^k(x^{\prime })r_k\right) ,$ }

{\normalsize $\hat{P}^{st}(x^{\prime })=\frac 1{\sqrt{2}}\left(
u_n^{st}(x^{\prime })-i\frac \delta {\delta u_{st}(x^{\prime
})}-N_n^{st^k}(x^{\prime })r_k\right) ,$ }

{\normalsize $A_{st}(x^{\prime })=\frac{\delta a^p}{\delta f_n^{st}(x)}%
\left( R_p^kr_k-iK_p\right) ,$ }

{\normalsize $A_n^{st}(x^{\prime })=\frac{\delta a^p}{\delta f_{st}(x)}%
\left( R_p^kr_k-iK_p\right) ,$ }

{\normalsize $K_p=S_p+\frac \partial {\partial \tau ^p},$ $r_k=R_k^pJ_p,$}

{\normalsize we have the same expressions for the operators of coordinate
and momentum of field operators: }

{\normalsize $\hat{q}_{st}=g\left( F_{st}(x^{\prime })+\frac 1g\hat{Q}%
_{st}(x^{\prime })+\frac 1{g^2}A_{st}(x^{\prime })\right) ,$ }

{\normalsize $\hat{p}^{st}=g\left( F_n^{st}(x^{\prime })+\frac 1g\hat{P}%
^{st}(x^{\prime })+\frac 1{g^2}A_n^{st}(x^{\prime })\right) .$ }

{\normalsize \ }

\section{ References:}

\subsection{General principles:}

\textrm{$[I]$ N.N.Bogoliubov, D.V.Shirkov INTRODUCTION to the THEORY 
of QUANTIZED FIELDS, (New-York - London, Interscience Publishes 1958)}

\textrm{$[II]$ C.W.Misner, K.S.Thorne, J.A.Wheeler GRAVITAION. (W.H.Freeman
and Company, San Francisco, 1973)}

\textrm{$[III]$ N.D.Birrell, P.C.W.Davies QUANTUM FIELD in the  
 CURVED SPACE (Cambrige University Press, 1982) }

\textrm{$[IV]$ R.Rajaraman. SOLITONS and INSTANTONS: an Introduction to 
Solitons and Instantins in Quantum Field Theory (North-Holland, 1982)}

\textrm{\ }

\subsection{Bogoliubov Transformation:}

\textrm{$[1]$ N.N.Bogoliubov // the Ukrainian Mathematical Journal. 1950. '.
2. '. 3-24. }

\textrm{$[2]$ E.P.Solodovnikova, A.N.Tavkhelidze, O.A.Khrustalev //
Teor.Mat.Fiz. 1972. T.10. }

\textrm{. 162-181; T. 11. p. 317-330; 1973. T. 12. p. 164-178. }

\textrm{$[3]$ O.D.Timofeevskaya // Teor.Mat.Fiz 1983. T. 54. p. 464-468. }

\textrm{$[4]$ N.H.Christ, T.D.Lee // Phys. Rev. D. 1975. V. 12. P.
1606-1627. }

\textrm{$[5]$ E.Tomboulis // Phys. Rev. D. 1975. V. 12. P. 1678-1683. }

\textrm{$[6]$ M.Greutz // Phys. Rev. D. 1975. V. 12. P. 3126-3144. }

\textrm{$[7]$ K.A.Sveshnikov // Teor.Mat.Fiz. 1985. T. 55. p. 361-384; 1988.
T. 74. '. }

\textrm{$[8]$ O.A.Khrustalev, M.V.Tchitchikina // Teor.Mat.Fiz. 1997. T.
111. N2.p. 242-251; }

\subsection{Gravity and Quantum Gravity:}

\textrm{$[9]$ R.Arnowitt, S.Deser, C.W.Misner//Phys. Rev 1960 V.120 N.1
p.863-870}

\textrm{$[10]$ E.T.Newman, L.Tamburino, T.Unti//J. Math. Phys. 1963 V.7 N.5
p 915-923}

\textrm{$[11]$ E.T.Newman, L.Penrose//J. Math. Phys. 1966 V.7 N.5 p.863-870}

\textrm{$[12]$ W.Misner//J. Math. Phys. 1967 V.4 N.7 p.924-937}

\textrm{$[13]$ Y.Choquet-Bruhat, R.Georch//Commun. Math. Phys. 1969 N.14
p.329-335}

\textrm{$[14]$ J.W.York//Phys. Rev. Letters 1971 V.26 N.26 p.1656-1658}

\textrm{$[15]$ J.W.York//J. Math. Phys. 1972 V.13 N.2 p.125-130}

\textrm{$[16]$ A.E.Fisher, J.E.Marsden // J. Math. Phys. 1972. V.13 N.4 p.
546-568 }

\textrm{$[17]$ K.Kuchar//J. Math. Phys. 1972 V.13 N.5 p.768-781}

\textrm{$[18]$ S.A.Fulling // Phys. Rev D 1973. V.7 N.10 p. 2850-2862}

\textrm{$[19]$ N.O.Murchadha, J.W.Jork // Phys. Rev D 1974. V.10 N.2 p.
428-446}

\textrm{$[20]$ T.Regge, C.Teiteloim // Annals of Phys. 1974. N.88 p.286-318 }

\textrm{$[21]$ D.Cramer // ACTA\ PHYSICA POLONICA 1975. N.4 p.467-478}

\textrm{$[22]$ P.Cordero, C.Teiteloim //Annals of Phys. 1976. N.100 p.607-631%
}

\textrm{$[23]$ A.E.Fisher, J.E.Marsden//Gen. Relat. and Grav. 1976 V.7 N.12
p.915-920}

\textrm{$[24]$ K.Kuchar//J. Math. Phys. 1976 V.17 N.5 p.777-780}

\textrm{$[25]$ G.W.Gibbsons, D.N.Page, C.N.Pope // Commun. Math.Phys. 1990.
N127 p. 529-553 }

\textrm{$[26]$ R.Gomes, P.Laguna, Ph.Papadopoulos, J.Winicour//
gr-qc/9603060 1996 }

\textrm{$[27]$ C.Rovelli// gr-qc/0006061 2000 V.2 }

$\left[ 28\right] $ O.A.Khrustalev, M.V.Tchitchikina // gr-qc/0109067 2001

\end{document}